\documentclass[12pt]{article}
\usepackage{ucs}
\usepackage[utf8x]{inputenc}
\usepackage{braket}
\usepackage{dsfont}
\usepackage{xcolor}
\usepackage[linktocpage]{hyperref}
\hypersetup{
    colorlinks = true,
    linkcolor=blue,
    citecolor=red,
    urlcolor=blue,
}
\usepackage{amscd,amsmath,amssymb}
\usepackage{graphics,graphicx}
\usepackage[mathscr]{euscript}
\def\a{\alpha}
\def\b{\beta}

\def\d{\delta}
\def\e{\epsilon}

\def\g{\gamma}

\def\r{\rho}
\def\s{\sigma}

\def\ve{\varepsilon}

\def\be{\begin{equation}}
\def\ee{\end{equation}}
\def\arr{\begin{array}{rll}}
\def\ea{\end{array}}
\def\bea{\begin{eqnarray}}
\def\eea{\end{eqnarray}}

\def\N2{$N{=}2$}

\def\>{\rangle}
\def\<{\langle}
\def\+{\dagger}
\def\={\ =\ }

\begin{document}

\title{
\hfill\\
\hfill\\
Three-dimensional (higher-spin) gravities with extended Schr\"odinger and $l$-conformal Galilean symmetries
\\[0.5cm]
}

\author{Dmitry Chernyavsky$\,{}^{a, b}$ and Dmitri Sorokin$\,{}^{c,d}$}

\date{}

\maketitle
\vspace{-1.5cm}

\begin{center}

\vspace{0.5cm}\textit{\small
${}^a$ School of Physics, Tomsk Polytechnic University,\\
634050 Tomsk, Lenin Ave. 30, Russia
\centerline {$\&$}
${}^b$ Tomsk State University of Control Systems and Radioelectronics,\\
634050 Tomsk, Lenin Ave. 40, Russia
}

\vspace{0.5cm}
\textit{\small
${}^c$ I.N.F.N., Sezione di Padova \\
\centerline {$\&$}
${}^d$ Dipartimento di Fisica e Astronomia ``Galileo Galilei",  Universit\`a degli Studi di Padova, \\
Via F. Marzolo 8, 35131 Padova, Italy
}
\end{center}
\vspace{5pt}

\abstract{We show that an extended $3D$ Schr\"odinger algebra introduced in \cite{Hartong:2016yrf} can be reformulated as a $3D$ Poincar\'e algebra extended with an SO(2) R-symmetry generator and an $SO(2)$ doublet of bosonic spin-1/2 generators whose commutator closes on $3D$ translations and a central element. As such, a non-relativistic Chern-Simons theory based on the extended Schr\"odinger algebra studied in \cite{Hartong:2016yrf} can  be reinterpreted as a relativistic Chern-Simons theory. The latter can be obtained by a contraction of the $SU(1,2)\times SU(1,2)$ Chern-Simons theory with a non principal embedding of $SL(2,\mathbb R)$ into $SU(1,2)$. The non-relativisic Schr\"odinger gravity of \cite{Hartong:2016yrf} and its extended Poincar\'e gravity counterpart are obtained by choosing different asymptotic (boundary) conditions in the Chern-Simons theory. We also consider extensions of a class of so-called $l$-conformal Galilean algebras, which includes the Schr\"odinger algebra as its member with $l=1/2$, and construct Chern-Simons higher-spin gravities based on these algebras. }

\noindent

\thispagestyle{empty}


\newpage

\setcounter{footnote}{0}

\tableofcontents

\newpage
\section{Introduction}
Three-dimensional theories of gravity and their supersymmetric and higher-spin extensions have been under extensive study for several decades. A characteristic feature of a majority of these models is that they describe massless gauge fields of spin $s>0$ which do not propagate in the three-dimensional bulk. As a manifestation of this feature, these theories admit a description in terms of Chern-Simons actions for gauge fields valued in the adjoint representation of corresponding symmetry groups, as was first observed for the case of $3D$ supergravity \cite{Achucarro:1987vz}. In spite of having on-shell zero field strengths (or curvatures), these theories exhibit a rich structure on the boundary of $3D$ manifolds and give rise to a variety of holographic dualities.

A main activity has been in studying relativistic (higher-spin) gravity models in $AdS_3$ and Minkovski space backgrounds. However, non-relativistic gravity theories based on different extensions of the $3D$ Galilean group have also attracted attention, in particular, in relation to non-AdS holography and its condensed matter applications (see e.g. \cite{Prohazka:2017pkc} for a review and references). A Chern-Simons (CS) formulation of Galilean gravity was put forward in \cite{Papageorgiou:2009zc} and further generalized to a wider class of models in \cite{Hartong:2016yrf} including a conformal non-projectable Ho$\breve{\rm r}$ava–Lifshitz gravity associated with an extended Schr\"odinger algebra.

In the latter theory (dubbed Schr\"odinger gravity) the authors of \cite{Hartong:2016yrf} found solutions with $z=2$ Lifshitz geometries. On the other hand, $z=2$ Lifshitz solutions were also found within relativistic higher-spin CS theories based on $SL(N,\mathbb R)\times SL(N,\mathbb R)$ gauge groups \cite{Gary:2012ms,Gary:2014mca}. As was shown in \cite{Lei:2015ika} these solutions of the Chern-Simons  theory (build of connections) are not equivalent to  Lifshitz solutions in a metric-like theory. In the case of  \cite{Hartong:2016yrf} it was shown that their Lifshitz solutions do not have this problem, since the Newton–Cartan Chern–Simons theory is not a Lorentzian metric theory.

One of the aims of this note is to show that the CS theory based on the extended Schr\"odinger algebra can actually be reinterpreted in terms of a relativistic CS theory. The reason is that the extended Schr\"odinger algebra has an $sl(2,\mathbb R)\sim so(1,2)$ subalgebra. Other generators of the extended Schr\"odinger algebra transform under a vector or a spinor representation of $sl(2,\mathbb R)$, or are  $sl(2,\mathbb R)$ singlets. Thus, the algebra acquires a form similar to a centrally extended $\mathcal N=2$, $D=3$ Poincar\'e superalgebra, but with a doublet of commuting spinor generators. The $SO(2) \sim U(1)$ generator of $2d$ Galilean rotations becomes the R-symmetry generator of this ``bosonic supersymmetry" algebra.

Upon having rewritten the extended Schr\"odinger algebra in the relativistic form, one finds that it can be obtained by a contraction of an $su(1,2)\oplus sl(2,\mathbb R)\oplus so(2)$ algebra or as a contraction and truncation of an $su(1,2)\oplus su(1,2)$ algebra. The latter is one of the real forms of $sl(3)\oplus sl(3)$. Its difference with respect to the conventional real form $sl(3,\mathbb R)\oplus sl(3,\mathbb R)$ has been discussed in the context of Chern-Simons spin-3 gravity e.g. in \cite{Campoleoni:2010zq,Campoleoni:2011hg}.

The above observations point at a relation of  Schr\"odinger gravity to Chern-Simons constructions of $3D$ higher-spin theories in the following sense. It is well known (see e.g. \cite{Ammon:2011nk,Campoleoni:2011hg,Castro:2012bc}) that the physical content and asymptotic behavior of a theory described by an $SL(N)\times SL(N)$ Chern-Simons action depends on the choice of particular vacuum boundary conditions which in the relativistic case are related to the choice of the embedding of $SL(2,{\mathbb R})$ into $SL(N)$. In other words, one and the same Chern-Simons action may describe physically different theories. In this respect, the Schr\"rodinger gravity can be regarded as a specific choice of a non-relativistic vacuum associated with an embedding of the Galilean group into the extended Schr\"rodinger group or its expansion to the $SU(1,2)\times  SU(1,2)$, or even higher-rank group underlying a certain Chern-Simons action.

In the second part of this paper we will consider extensions of a class of so-called $l$-conformal Galilean algebras \cite{Negro:1997,Henkel:1997zz} which includes the Schr\"odinger algebra as its member with $l=1/2$, construct Chern-Simons higher-spin gravities based on these algebras (which turn out to be a subclass of so-called Hietarinta algebras \cite{Hietarinta:1975fu}) and discuss  asymptotic symmetries in these theories.

\section{Extended Schr\"odinger as extended Poincar\'e}
In this Section we will show that the extended Schr\"odinger algebra associated with a Galilean $d=2$ space can be recast in a relativistic form as an extended $D=2+1$ Poincar\'e algebra.
Our staring point is the $d=2$ Schr\"odinger algebra written in the standard basis
\begin{align}\label{Algebra}
&
[I,P^i]=\e^{ij}P^j\ , && [ I,G^i ]=\e^{ij}G^j , && [H,G^i]=P^i,
\nonumber\\[2pt]
&
[D,H]=-2H\, && [H,K]=D\ , &&[D,K]=2K,
\nonumber\\[2pt]
&
[H,G^i]=P^i\, && [D,P^i]=-P^i,
\nonumber\\[2pt]
&
[D,G^i]=G^i\, && [K,P^i]=-G^i,
\end{align}
where $H$, $K$ and $D$ are, respectively, the generators of time translations, special conformal transformations and dilatations forming the one-dimensional conformal algebra isomorphic to $sl(2,\mathbb R)$. $P^i$ and $G^i$ ($i=1,2$) generate spatial translations and Galilei boosts,  while $I$ generates the $SO(2)$ rotations in the $2d$ Galilean space.
It is known that the commutator of translations and Galilean boosts can be centrally extended $[P^i,G^j]=N\d^{ij}$ and, when one considers the Galilean algebra only, the result is the so called Bargmann algebra. In \cite{Hartong:2016yrf} it was proposed to extend the  Scr\"odinger algebra further by adding three new elements which appear in the commutators of the Galilean boosts and translations
\be\label{Commutators_Translations and Boosts}
[G^i,G^j]=S\e^{ij}, \qquad [P^i,P^j]=Z\e^{ij},\qquad [P^i,G^j]=N\d^{ij}-Y\e^{ij}.
\ee
The new elements  $S$, $Y$ and $Z$ are central with respect to the Galilean subalgebra, but have nontrivial commutation relations with the conformal subalgebra generators
\begin{align}\label{HPoincare}
&
[H,Y]=-Z\,, && [H,S]=-2Y\,, && [K,Y]=S,
\nonumber\\[2pt]
&
[K,Z]=2Y\,, && [D,S]=2S\,, && [D,Z]=-2Z.
\end{align}
It turns out that the above commutation relations form a $D=2+1$ Poincar\'e algebra. In order to see this, let us redefine the generators as follows
\bea
&&
Z=M_{-1},\qquad S=M_{+1},\qquad Y=M_0,
\nonumber\\[2pt]
&&
H=L_{-1}, \qquad K=L_{+1}, \qquad D=-2L_0.
\eea
Upon this redefinition the algebra (\ref{HPoincare}) take the form of the $D=2,1$ Poincar\'e algebra written in the $BMS_3$ basis
\be\label{Algebra_Poincare}
[L_m,L_n]=(m-n)L_{m+n},\qquad [L_m,M_n]=(m-n)M_{m+n}, \qquad m, n=\pm1,0.
\ee
As the next step let us combine the Galilean translations and boosts into a single set of generators $Z^i_{\alpha}$ $(\alpha=\mp \frac 12)$ such that
\be\label{Redefinitio_Translations to Spinor}
 P^i=-\sqrt{2}Z^i_{-\frac12}, \qquad G^i=\sqrt{2} Z^i_{+\frac12}
\ee
Now we can rewrite the commutation relations (\ref{Algebra}) and (\ref{Commutators_Translations and Boosts}) as follows
\bea\label{Algebra_Infinite-like basis}
&&
[L_m,Z^i_\alpha]=\left(\frac{m}{2}-\alpha\right)Z^i_{m+\alpha}, \qquad [I,Z^i_\alpha]=\e^{ij}Z^j_\alpha,
\nonumber\\[2pt]
&&
[Z^i_\alpha,Z^j_\beta]=\frac 12\e^{ij}M_{\alpha+\beta}+\frac 12 N(\alpha-\beta)\d^{ij}, \qquad m,n=\pm1,0, \quad \alpha,\beta=\pm\frac12.
\eea
Curiously, the structure of these relations resembles the form of a centrally extended $\mathcal{N}=2$, $D=3$ Poincar\'e superalgebra, 
but with the commuting (bosonic) spinor generators $Z^i_\alpha$ instead of anti-commuting ones. To make this similarity more explicit, let us now rewrite the extended Schr\"odinger algebra in a manifestly $D=2+1$ Lorentz invariant form. To this end let us perform the following redefinition
\begin{align}\label{Redefinition_To Lorentz}
&
\sqrt2 J^-=-L_{-1}=H\ , && \sqrt2 {\mathcal P}^-=-M_{-1}=-Z\ , 
\nonumber\\[2pt]
&
\sqrt2 J^+=L_{+1}=K\ , &&  \sqrt2 {\mathcal P}^+=M_{+1}=S\ ,  
\nonumber\\[2pt]
&
J^2=L_0=-\frac 12D\ , && {\mathcal P}^2=M_0=Y.
\end{align}
Then the algebra (\ref{Algebra_Poincare}) and (\ref{Algebra_Infinite-like basis}) take the form of a $3D$ relativistic algebra \footnote{See Appendix A for our notation and conventions.}
\bea\label{Algebra_Lorentz-like}
&&
[J^a,J^b]=\e^{abc}J_c, \qquad [J^a, {\mathcal P}^b]=\e^{abc} {\mathcal P}_c, \qquad [I,Z^i_\a]=\e^{ij} Z^j_\a,
\nonumber\\[2pt]
&&
[J^a,Z^i_\b]=\frac12 Z^i_\b (\gamma^a)^\b{}_\a, \qquad [Z^i_\a, Z^j_\b]=-\frac12 \e^{ij}(C\gamma_a)_{\a\b}{\mathcal P}^a+N C_{\a\b}\d^{ij},
\eea
where we have also rescaled $N\rightarrow -2N$. From the structure of \eqref{Algebra_Lorentz-like} we see that $J^a$ generate $SO(1,2)$ rotations and ${\mathcal P}^a$ generate $3D$ translations thus forming the $3D$ Poincar\'e algebra, while $Z^i_\a$ plays the role of the SO(2) doublet of $SL(2,\mathbb R)$-spinors generating a ``bosonic supersymmetry". The generator $I$ of the $2d$ Galilean rotations is now traded for the $SO(2)$ $R$-symmetry generator. We have thus shown that the extended Schr\"odinger algebra is isomorphic to an extended Poincar\'e one. As one could notice, upon passing from one form to another, the geometrical meaning of the generators change. In particular, the generator $I$ of the $2d$ Galilean rotations is now traded for the $SO(2)$ $R$-symmetry generator, the generators of the $1d$ conformal algebra become that of the $SO(1,2)$ rotations,  while the Galilean translations and boosts form the doublet of $SL(2,\mathbb R)$ spinors according to (\ref{Redefinitio_Translations to Spinor}) and (\ref{Redefinition_To Lorentz}).

\subsection{Gravity theory with extended Schr\"odinger symmetry}

The extended Schr\"odinger algebra has a nonsingular bilinear form. In \cite{Hartong:2016yrf} it was used  to construct a Chern-Simons action with the gauge group generated by (\ref{Algebra}) and (\ref{Commutators_Translations and Boosts}) and to derive in this way a novel version of the non--projectable conformal Ho$\breve{\rm r}$ava–Lifshitz gravity.
The isomorphism of the extended Schr\"odinger algebra and the extended Poincar\'e algebra established in the previous Section can be used to reformulate the Chern-Simons action of \cite{Hartong:2016yrf} as a relativistic model. In order to do so, let us present the non-degenerate symmetric bilinear form of the extended Schr\"odinger algebra in the relativistic  basis (\ref{Algebra_Lorentz-like})
\be\label{Bilinear Form}
\langle J^a,{\mathcal P}^b \rangle=\eta^{ab}, \qquad \langle Z^i_\a, Z^j_\b\rangle=C_{\a\b}\e^{ij}, \qquad \langle I,  N\rangle=-1.
\ee
The Chern-Simons action
\be\label{Action_CS}
S=\frac{k}{4\pi}\int_{\mathcal{M}_3}\langle \mathbf{A}d\mathbf{A}+\frac{2}{3}\mathbf{A}^3\rangle
\ee
is constructed with the use of a one--form gauge connection $\mathbf{A}=dx^\mu A_\mu(x)$ taking values in the algebra (\ref{Algebra_Lorentz-like}) and having the following components
\be
\mathbf{A}=e^a {\mathcal P}_a+\omega^a J_a+\lambda^{i\a} Z^i_\a+N v+I b.
\ee
In \eqref{Action_CS} the wedge-product of the differential forms is implied.
 To have a contact with the Einstein gravity the value of the level $k$ is set to be $k=1/(4 G)$ with $G$ being the Newton's constant.

Using the expression for the bilinear form (\ref{Bilinear Form}), the action (\ref{Action_CS}) can be rewritten (up to a boundary term) as follows
\be\label{Action_Extended Einstein}
S=\frac{k}{4\pi} \int_{\mathcal{M}_3}\left( 2 R_a\, e^a-\e^{ij} (\bar\lambda^i \nabla\lambda^j)-2v db\right),
\ee
where the covariant derivative is defined as
\be
\nabla\lambda^i=d\lambda^i+\frac12 \omega^a\gamma_a \lambda^i-b\e^{ij}\lambda^j
\ee
and
\be\label{R}
R^a=d\omega^a+\frac12 \e^{abc}\omega_b\omega_c
\ee
is the curvature associated with the $SL(2,\mathbb R)$ connection $\omega^a$.

The first term in \eqref{Action_Extended Einstein} is the action for Einstein gravity written in the first-order formalism with $e^a(x)$ being associated with the gravitational dreibein field.
As usual, the equations of motion of the Chern-Simons theory imply that the curvature  ${\mathbf F}_2=d{\mathbf A}+{\mathbf A}^2$ vanishes and locally $\mathbf A$ is a pure gauge, which implies that all the fields are non-dynamical in the $3D$ bulk and the non-trivial properties of the theory are determined by their behaviour on the $2d$ boundary. In this respect, the fact that the spin-3/2 fields $\lambda^i_\alpha$ have the bosonic statistics is not as troublesome as in higher dimensional theories, but still may indicate that unitarity may be lost on the boundary.  We will study this issue for the asymptotic symmetry group of this theory in Section \ref{ui}.

As usually, we can construct a relativistic metric tensor with the use of the fields $e^a$ and an affine connection (associated with $\omega^a$), whose antisymmetric part is defined by the torsion constructed with the fields $\lambda^i_\alpha$. A vacuum solution of the field equations for such a system is clearly the flat $3D$ Minkowski space.
On the other hand, as was considered in \cite{Hartong:2016yrf}, the Chern-Simons action based on the extended Schr\"odinger algebra allows one to define another metric and affine connection which correspond to a non-relativistic $3D$ geometry. In our notation, the metric of the Galilean geometry was constructed in \cite{Hartong:2016yrf} with the one-forms  $\omega^0$ and $\lambda^i_{+\frac 12}$ (associated, respectively, with the generators $H$ and $P^i$ in \eqref{Redefinition_To Lorentz} and  \eqref{Redefinitio_Translations to Spinor}) which play the role of the Galilean dreibein. Evidently, this corresponds to a different choice of geometry and boundary conditions for the Chern-Simons field equations of motion. These alternative choices result in physically different theories. The situation is analogous to different choices of the embedding of the $SL(2,\mathbb R)$ group into $SL(N)\times SL(N)$ Chern-Simons theories which (together with asymptotic boundary conditions) lead to different $3D$ higher-spin gravity models (see e.g. \cite{Ammon:2011nk,Campoleoni:2011hg,Castro:2012bc,Afshar:2012hc,Bunster:2014mua} and references therein).

Another curious fact about the action  (\ref{Action_Extended Einstein}) is that it can be obtained in a limit of zero cosmological constant from the $SU(1,2)\times SU(1,2)$ Chern-Simons theory. Or putting it differently, the extended Schr\"odinger gravity can be expanded to the latter.

\subsection{The extended Schr\"odinger gravity by contraction and truncation of $SU(1,2)\times SU(1,2)$ Chern-Simons theory}
As in the case of its $sl(3,\mathbb R)$ counterpart \cite{Ammon:2011nk}, the $su(1,2)$ algebra allows for two $sl(2,\mathbb R)$ embeddings, the principle embedding and a non-principle one (see Appendix B). It turns out that the extended Schrodinger algebra
is related to the non-principle embedding for which the $su(1,2)$ algebra takes the following form
\bea\label{Algebra_AdS_l=1/2}
&&
[\mathcal{J}^a,\mathcal{J}^b]=\e^{abc}\mathcal{J}_c, \qquad \qquad \hspace{1mm} [\mathcal{I}, \mathcal{Z}^i_\a]=\e^{ij}\mathcal{Z}^j_\a,
\nonumber\\[2pt]
&&
[\mathcal{J}^a, \mathcal{Z}^i_\a]=\frac12 (\g^a)^\b{}_\a \mathcal{Z}^i_\b, \qquad [\mathcal{Z}^i_\a, \mathcal{Z}^j_\b]=-\frac12 \e^{ij} (C\g_a)_{\a\b} \mathcal{J}^a+\frac34 \d^{ij} C_{\a\b}\mathcal{I}\,,
\eea
where $J^a$ form the $sl(2,\mathbb R)$ subalgebra.

Let us now consider two copies of (\ref{Algebra_AdS_l=1/2}) which form the $su(1,2)\oplus su(1,2)$ algebra, and take the following linear combination of their generators which are distinguished by `tilde'
\bea\label{Contraction}
&&
{\mathcal P}^a=\frac{1}{\rho} (\mathcal{J}^a-\tilde{\mathcal{J}}^a), \qquad J^a=\mathcal{J}^a+\tilde{\mathcal{J}}^a, \qquad Z^i_{\a}=\sqrt{\frac{2}{\rho}} \mathcal{Z}^{i}_{\a}, \qquad  \tilde Z^i_{\a}=\sqrt{\frac{2}{\rho}} \tilde{\mathcal{Z}}^{i}_{\a}\,,
\nonumber\\[2pt]
&&
I=\mathcal{I}-\tilde{\mathcal{I}}, \qquad N=\frac{3}{4\rho}(\mathcal{I}+\tilde{\mathcal{I}}).
\eea
The generators $\mathcal P^a$ and $J^a$ form the $so(2,2)$ algebra of $AdS_3$ isometry, and  $\rho$ can be viewed as the $AdS_3$ radius. When $\rho\to \infty$, the $so(2,2)$ algebra gets contracted to the $3D$ Poincar\'e algebra.
Taking also into consideration the generators $Z^i_{\a}$, $\tilde Z^i_{\a}$, $I$ and $N$, in the limit $\rho\rightarrow \infty$ one recovers the extended Schr\"odinger algebra in the form (\ref{Algebra_Lorentz-like}) but with the extra doublet  $\tilde{Z}^i_{\a}$ of the spinor generators
\be\label{Z-}
[\tilde{Z}^i_{\a},\tilde{Z}^j_{\b}]=\frac12 \e^{ij} (C\g_a)_{\a\b} \mathcal{P}^a+ \d^{ij} C_{\a\b}N\,.
\ee
We see that the generators $\tilde{Z}^i_{\a}$ further extend the algebra (\ref{Algebra_Lorentz-like}). In the non-relativistic setting, these generators correspond to an extra copy of the Galilei-like translations and boosts $\tilde{Z}^i_{\a}=(\tilde{P}^i,\tilde{G}^i)$. \footnote{\label{2}  It might be of interest to see whether the extended Schr\"odinger algebra (\ref{Algebra_Lorentz-like}) with the addition of \eqref{Z-} can be alternatively viewed as a certain algebra expansion, a technique considered e.g. in \cite{Khasanov:2011jr,Hansen:2019vqf,Bergshoeff:2019ctr} and references there in. } The extended Schr\"odinger algebra is obtained upont trancation of these additional generators. An alternative possibility, which does not require the truncation of extra spinor generators, is to obtain the extended Schr\"odinger algebra directly by contraction of $su(1,2)\oplus sl(2,{\mathbb R}) \oplus so(2)$.\footnote{Note that, instead of the contraction of $su(1,2)\oplus su(1,2)$ we could also consider the contraction of $sl(3,R)\oplus sl(3,R)$ by simply assuming that the vector indices $i,j$ in  (\ref{Action_AdS}) be transformed under the $SO(1,1)$ group instead of $SO(2)$ (see Appendix B). However, in that case, because of non-compactness of $SO(1,1)$ we would arrive at an algebra which would not have an interpretation as an extended Schr\"odinger (or Galilean) algebra. From the point of
view of a non-relativistic gravity interpretation a somewhat similar case in which one
deals with a different (non-compact) real form is dubbed pseudo-Newton–Cartan
geometry  \cite{Hartong:2017bwq}. We are thankful to a referee for indicating this and the point of footnote \ref{2}.}

The above observation of the relation between the algebras allows us to view the  gravity model (\ref{Action_Extended Einstein}) as a contraction and truncation of the $SU(1,2)\times SU(1,2)$ CS theory \cite{Campoleoni:2010zq}. For our case of the non-principle embedding of $sl(2,\mathbb R)$, it is natural to define the $SU(1,2)\times SU(1,2)$ gauge field one--form $\mathbf{A}$ in the basis (\ref{Contraction})
\be
\mathbf{A}=\mathbf{A}+\mathbf{\tilde A}=e^a P_a+\omega^a J_a+\lambda^{i,\a} Z^i_{\a}+\tilde\lambda^{i,\a} \tilde Z^i_{\a}+N v+I b.
\ee
Then, using the invariant bilinear forms\footnote{Using the  redefinitions (\ref{Contraction}), (\ref{Redefinition_su(1,2)_1}) and (\ref{Redefinition_su(1,2)_2}) one can see that (up to a normalization constant) the bilinear forms (\ref{Bilinear_su(1,2)+su(1,2)}),  correspond to the difference between the two copies of the bilinears (\ref{Bilinear_su(1,2)}) of the $su(1,2)$ algebra. In other words, the CS action \eqref{Action_AdS} is equal to $\frac k{4\pi}\int_{{\mathcal M}_3}\left ( \langle \mathbf{A}d\mathbf{A}+\frac{2}{3}\mathbf{A}^3\rangle-\langle \mathbf{\tilde A}d\mathbf{\tilde A}+\frac{2}{3}\mathbf{\tilde A}^3\rangle\right)$.}
\be\label{Bilinear_su(1,2)+su(1,2)}
\langle J^a,P^b \rangle=\eta^{ab}, \quad \langle I,  N\rangle=-1,
\quad \langle Z^i_\a, Z^j_\b\rangle=C_{\a\b}\e^{ij},\quad \langle \tilde Z^i_\a, \tilde Z^j_\b\rangle=-C_{\a\b}\e^{ij},
\ee
one gets the $SU(1,2)\times SU(1,2)$ CS action in the form
\be\label{Action_AdS}
S=\frac{k}{4\pi}\int_{\mathcal{M}_3} \left( 2 e^a\, R_a+\frac{1}{3\rho^2}\e_{abc}e^a e^b e^c-\e^{ij}\bar{\lambda}^i \nabla\lambda^j +\e^{ij}\overline{\tilde \lambda}^i \nabla\tilde \lambda^j-2v db\right),
\ee
where
\be
\nabla\lambda^i=d\lambda^i+\frac12 \omega^a\gamma_a\lambda^i+\frac{1}{2\rho}e^a\gamma_a \lambda^i-\frac{3}{4\rho}\e^{ik}v\lambda^k-\e^{ik}b\lambda^k
\ee
and
$$
\tilde\nabla\tilde\lambda^i=d\tilde\lambda^i+\frac12 \omega^a\gamma_a\tilde\lambda^i-\frac{1}{2\rho}e^a\gamma_a \tilde\lambda^i-\frac{3}{4\rho}\e^{ik}v\tilde\lambda^k+\e^{ik}b\tilde\lambda^k.
$$
We see that in the form (\ref{Action_AdS}) the CS action describes gravity (associated with the dreibein $e^a$ and spin connection $\omega^a$) plus bosonic spin-3/2 fields $\lambda^{i}_\alpha$ and $\tilde\lambda^i_\alpha$ coupled to gravity and a pair of spin-1 fields $v$ and $b$, which is known to be the consequence of the choice of the non-principle embedding of $SL(2,\mathbb R)$ into $SU(1,2)$.
The action (\ref{Action_Extended Einstein}) is obtained from (\ref{Action_AdS}) by taking the limit $\rho\rightarrow\infty$  and setting the fields $\tilde\lambda^i$ to zero. This is consistent with the field variations under the local symmetries generated by the extended Schr\"odinger algebra (\ref{Algebra_Lorentz-like}). As one might expect, the theory obtained in this limit is different from the usual asymptotically flat spin-3 gravity discussed e.g. in \cite{Afshar:2013vka,Gonzalez:2013oaa,Matulich:2014hea}.

\subsection{Asymptotic symmetry}\label{as}
Let us now analyze asymptotic symmetry of the gravity theory \eqref{Action_Extended Einstein} based on the extended Schr\"odinger group assuming that the $3D$ geometry is relativistic and described by the dreibein $e^a$ and the connection $\omega^a$.

As we have shown in the previous section this theory can be obtained by contraction and truncation of the $SU(1,2)\times SU(1,2)$ CS theory with a non-principle embedding of $SL(2,\mathbb R)$ into $SU(1,2)$. One can thus expect that the asymptotic symmetry of \eqref{Action_Extended Einstein} can be recovered by a contraction of the asymptotic symmetry of the $SU(1,2)\times SU(1,2)$ CS theory.
In the case of the principle embedding the asymptotic symmetry of the latter was identified  with a $W_3\times W_3$ algebra in \cite{Campoleoni:2010zq}. In the case of the non-principle embedding of $SL(2,\mathbb R)$ in the  $SL(3,\mathbb R)\times SL(3,\mathbb R)$ theory it was shown \cite{Ammon:2011nk} that the asymptotic algebra is the direct product of two copies of a $W_3^{(2)}$ algebra (also known as the Bershadsky-Polyakov algebra \cite{Bershadsky:1990bg,Polyakov:1989dm}). In the  $SU(1,2)\times SU(1,2)$ case the asymptotic algebra is a different real form of $W_3^{(2)}$, which we will call $W_{1,2}^{(2)}$.

In the Appendix D we will obtain the asymptotic symmetry algebra of the theory \eqref{Action_Extended Einstein} by contraction and truncation of the $W_{1,2}^{(2)} \oplus W_{1,2}^{(2)}$ algebra, while in this Section we derive it directly.

We assume that the boundary of the $3D$ manifold $\mathcal{M}_3$ has a cylindrical topology with the compact directions parameterized by the coordinate $\phi$ and the non-compact one is $t$. The radial coordinate $r$ measures how far we are from the boundary. As usually, we assume that at the boundary the gauge field behaves  as
\be\label{Boundary gauge field}
\mathbf{A}=h^{-1}(d+\mathfrak{a})h,
\ee
where a group element $h$ depends on the radial coordinate $h=h(r)$ only. When all the fields, except the ones defining Einstein gravity are set to zero, one assumes to have the $BMS_3$  boundary conditions (see e.g. \cite{Ashtekar:1996cd,Barnich:2006av,Barnich:2014cwa})
\bea\label{Boundary_Conditions_BMS}
&&
\mathfrak{a}_\phi^{0}=-\mathcal{L} M_{-1}-\mathcal{M}L_{-1}+L_{+1},
\nonumber\\[2pt]
&&
\mathfrak{a}_t^{0}=-\mathcal{M}M_{-1}+M_{+1},
\eea
where we used the basis \eqref{Algebra_Poincare} and \eqref{Algebra_Infinite-like basis} of the generators of the extended Schr\"odinger algebra.
As an extension of \eqref{Boundary_Conditions_BMS} we define the following boundary conditions for the connection $\mathfrak{a}$
\bea\label{Boundary_Conditions}
&&
\mathfrak{a}_\phi=\mathfrak{a}_\phi^{0}+\mathcal{N} \mathcal{I} M_{-1}+\mathcal{N}^2 L_{-1}+\sqrt2\mathcal{C}^i Z^i_{-\frac12}+\mathcal{I}N+\mathcal{N} I,
\nonumber\\[2pt]
&&
\mathfrak{a}_t=\mathfrak{a}_t^{0}+\mathcal{N}^2M_{-1}+2\mathcal{N} N,
\eea
where $\mathcal{M}$, $\mathcal{N}$, $\mathcal{I}$, $\mathcal{L}$ and $\mathcal{C}$ are functions of $\phi$ and $t$ describing asymptotic dynamics of the fields of the model. Specifying appropriately the element $h$, these boundary conditions include physically interesting solutions, such as cosmological horizons \cite{Bagchi:2012xr,Barnich:2012aw}. One can notice a close relation of these boundary conditions to the ones proposed in \cite{Fuentealba:2017fck} for asymptotically flat $\mathcal{N}=2$, $D=3$ supergravity. It can be verified that the gauge field  (\ref{Boundary_Conditions}) satisfies the equations of motion $d\mathbf{A}+\mathbf{A}^2=0$, provided
\be\label{Equations of motion_Solution}
\dot{\mathcal{L}}=\mathcal{M}', \qquad \dot{\mathcal{I}}=2\mathcal{N}', \qquad \dot{\mathcal{N}}=\dot{\mathcal{M}}=\dot{\mathcal{C}}^i=0,
\ee
where, hereinafter, prime and dot denote, respectively, the derivative with respect to $\phi$ and $t$. The boundary conditions (\ref{Boundary_Conditions}) thus ensure a well defined variation principal of the CS action (\ref{Action_CS}) for getting the equations of motion such that
\be
\d S|_{EOM}=\frac{k}{4\pi}\int_{\partial \mathcal{M}} \langle \d \mathbf{A}, \mathbf{A}\rangle=0.
\ee
Indeed, taking into account the bilinear form (\ref{Bilinear Form}) written in the basis (\ref{Algebra_Poincare}) and (\ref{Algebra_Infinite-like basis})
\be\label{Bilinear Form_BMS-like}
\langle L_{+1},M_{-1} \rangle=\langle L_{-1},M_{+1} \rangle=-\langle I, N \rangle=-2, \quad \langle L_{0},M_{0} \rangle=1, \quad \langle Z^i_{-\frac12}, Z^j_{+\frac12} \rangle=\e^{ij},
\ee
one can explicitly check that the integrand in the expression above vanishes for the boundary conditions (\ref{Boundary_Conditions}).

 We now look for the transformations
 \be\label{Transformation}
 \d \mathbf{A}=d\boldsymbol{\lambda}+[\mathbf{A}, \boldsymbol{\lambda}],
  \ee
  that preserve  (\ref{Boundary_Conditions}), i.e. the transformations which map the boundary conditions to the same class.  We find that the algebra-valued parameter $\boldsymbol{\lambda}$ of these transformations should be of the following form
\bea\label{lambda_1/2}
&&
\boldsymbol{\lambda}=\left(\frac12 \ve_L''-\ve_L(\mathcal{M}-\mathcal{N}^2)\right)L_{-1}
\nonumber\\[2pt]
&&
+\left(\frac12 \ve_M''-\ve_L(\mathcal{L}-\mathcal{N}\mathcal{I})-\ve_M(\mathcal{M}-\mathcal{N}^2)-\frac12 \mathcal{C}^i \ve^i\right) M_{-1}
\nonumber\\[2pt]
&&
+\sqrt2 (\ve'^j\e^{ij}+\ve_L \mathcal{C}^i+\mathcal{N}\ve^i)Z^{i}_{-\frac12}+(\ve_I+\mathcal{N}\ve_L)I+(\ve_N+2\ve_M \mathcal{N}+\ve_L \mathcal{I})N
\nonumber\\[2pt]
&&
-\ve_L'L_0+\ve_L L_{+1}-\ve'_M M_{0}+\ve_{M}M_{+1}-\sqrt2\e^{ij}\ve^j Z^i_{+\frac12},
\eea
where the parameters $\ve$ are subject to the constraints
\be\label{Parameters_Constraint}
\dot{\ve}_M=\ve'_L, \qquad \dot{\ve}_N=2\ve'_I, \qquad \dot{\ve}_L=\dot{\ve}^i=\dot{\ve}_I=0,
\ee
and depend arbitrarily on $\phi$.
The transformations generated by $\boldsymbol{\lambda}$ imply the following transformation rules for the functions describing boundary dynamics
\bea\label{Transformation rules}
&&
\d \mathcal{L}=2\ve_L' \mathcal{L}+\ve_L \mathcal{L}'+2\ve_M' \mathcal{M}+\ve_M \mathcal{M}'+\frac32  \mathcal{C}^i\ve'^i+\frac12\mathcal{C}'^i \ve^i+\mathcal{I}\ve_I'+\mathcal{N}\ve_N'-\frac12 \ve_M''',
\nonumber\\[2pt]
&&
\d \mathcal{M}=2\ve_L' \mathcal{M}+\ve_L \mathcal{M}'+2\ve_I' \mathcal{N}-\frac12 \ve_L''',
\nonumber\\[2pt]
&&
\d \mathcal{C}^i=\frac32 \ve_L' \mathcal{C}^i+\ve_L \mathcal{C}'^i-\mathcal{M}\e^{ij}\ve^j+2\ve'^i \mathcal{N}+\ve^i \mathcal{N}'+\e^{ij}\mathcal{C}^j\ve_I+\e^{ij}\ve''^j,
\nonumber\\[2pt]
&&
\d\mathcal{N}=\ve'_I+\mathcal{N}'\ve_L+\mathcal{N}\ve_L',
\nonumber\\[2pt]
&&
\d\mathcal{I}=\ve_N'+2(\ve_M' \mathcal{N}+\ve_M\mathcal{N}')+\ve_L' \mathcal{I}+\ve_L \mathcal{I}'+\e^{ij}\mathcal{C}^i\ve^j.
\eea
For each symmetry transformation there is an associated conserved charge $Q[\boldsymbol{\lambda}]$ whose field variation in the CS theory is (see e.g. \cite{Banados:1994tn} for details)
\be\label{Charge variation}
\d Q[\boldsymbol{\lambda}]=-\frac{k}{2\pi}\int_0^{2\pi} \langle \boldsymbol{\lambda}, \d \mathfrak{a}_\phi \rangle d\phi.
\ee
Taking into account the bilinear form (\ref{Bilinear Form_BMS-like}) and the expression for $\boldsymbol{\lambda}$ (\ref{lambda_1/2}), one finds
\be\label{Bilinear_Calc}
\langle \boldsymbol{\lambda}, \d \mathfrak{a}_\phi \rangle=-2\left(\ve_I \d \mathcal{I}+\ve_N \d \mathcal{N}+\ve_M\d \mathcal{M}+\ve_L \d \mathcal{L}+\ve^i \d \mathcal{C}^i\right).
\ee
The variation (\ref{Charge variation}) defines the Poisson bracket of the charges
$$\d_{\boldsymbol{\lambda}_2}Q[\boldsymbol{\lambda}_1]={\Big[}Q[\boldsymbol{\lambda}_1], Q[\boldsymbol{\lambda}_2]{\Big]},$$
which form the classical algebra of the asymptotic symmetries.
To  get an explicit form of this algebra, it is common to expand the fields and the transformation parameters in Fourier modes. In view of (\ref{Charge variation}) and (\ref{Bilinear_Calc}), the Fourier modes of the conserved charges are given by
$Q_n=\frac{k}{\pi}\int_0^{2\pi} d\phi e^{-in\phi} X$, where $X$ stands for the functions, describing asymptotic dynamics. Using this expression together with the transformation rules (\ref{Transformation rules}), one finds the asymptotic symmetry algebra
\bea\label{Asymptotic_Algebra}
&&
[L_m,L_n]=i(m-n)L_{m+n}, \qquad \quad [L_m,M_n]=i(m-n)M_{m+n}-i k n^3\d_{m+n,0},
\nonumber\\[2pt]
&&
[L_m, I_n]=-i n I_{m+n}, \qquad \qquad \hspace{6.5mm} [L_m, N_n]=-i n N_{m+n},
\nonumber\\[2pt]
&&
[M_m, I_n]=-2 i n N_{m+n}, \qquad \qquad \hspace{2mm} [I_m, N_n]=-2 i n k \d_{m+n,0},
\nonumber\\[2pt]
&&
[L_m,C^i_p]=i\left(\frac{m}{2}-p\right)C_{p+m}^i, \qquad \hspace{1mm} [I_m, C^i_p]=-\e^{ij}C^j_{m+p},
\nonumber\\[2pt]
&&
[C_p^i, C_q^j]=-M_{p+q}\e^{ij}+i(p-q) N_{p+q} \d^{ij}-2 q^2 k \d_{p+q,0} \e^{ij},
\eea
where $m, n \in \mathbb{Z}$ and $p, q \in \mathbb{Z}+\frac12$. The generators are associated to the asymptotic fields as follows $(L, M, I,$ $N, C)\sim (\mathcal{L}, \mathcal{M}, \mathcal{I},\mathcal{N}, \mathcal{C})$. The first line represents the $BMS_3$ algebra with the standard central charge $c=12 k$ \cite{Barnich:2006av}.\footnote{Infinite dimensional extensions of the extended Schr\"odinger algebra and their connection with the $BMS_3$ algebra were earlier noticed by Yang Lei (unpublished). We thank Jelle Hartnog for pointing this out to us.} As one might expect, the algebra has a very similar form to that of the asymptotic symmetry superalgebra in an $\mathcal{N}=2$, $D=3$ supergravity theory \cite{Fuentealba:2017fck}.

\subsection{Unitarity issue}\label{ui}
The theory at the boundary is governed by representations of the asymptotic symmetry algebra (\ref{Asymptotic_Algebra}). An important question is whether the algebra under consideration has unitary representations. Note that though the unitarity issues with the $W^{(2)}_3\oplus W^{(2)}_3$ algebra are known \cite{Castro:2012bc,Afshar:2012hc}, the existence of unitary representations of the contraction of these algebra may not be excluded {\it a priori}.

Unitarity implies that one must have a positive semi--definite Kac matrix, which is a matrix constructed out of inner products between descendents at level $N$. We will define the notion of level and descendents for the algebra below. Here we aim to explore the Kac matrix at the first two levels and derive constraints imposed by the positivity condition.

To proceed, let us redefine the generator $M_m\rightarrow M_m - \frac{k}{2} \d_{m,0}$ and replace the Dirac brackets with the quantum commutators $[,]\rightarrow i[,]$ to bring the algebra (\ref{Asymptotic_Algebra}) to the following (quantum) form
\bea\label{Asymptotic_Algebra_Unitarity}
&&
[L_m,L_n]=(m-n)L_{m+n}, \qquad \quad [L_m,M_n]=(m-n)M_{m+n} - k (n^3-n)\d_{m+n,0},
\nonumber\\[2pt]
&&
[L_m, I_n]=-n I_{m+n}, \qquad \qquad \hspace{6.5mm}[L_m, N_n]=- n N_{m+n},
\nonumber\\[2pt]
&&
[M_m, I_n]=-2 n N_{m+n}, \qquad \qquad \hspace{2mm} [I_m, N_n]=-2 n k \d_{m+n,0},
\nonumber\\[2pt]
&&
[L_m,C^i_p]=\left(\frac{m}{2}-p\right)C_{p+m}^i,  \qquad\hspace{2mm} [I_m, C^i_p]=i\e^{ij}C^j_{m+p},
\nonumber\\[2pt]
&&
[C_p^i, C_q^j]=iM_{p+q}\e^{ij}+(p-q) N_{p+q} \d^{ij}+2 i k \left(q^2-\frac14\right) \d_{p+q,0} \e^{ij}.
\eea
As usual, we consider the generators of this algebra as operators acting on a vector space of quantum states. The operators have the hermiticity relations
\bea
&&
L_m^\dag=L_{-m}, \qquad M_m^\dag=M_{-m}, \qquad (C^i_p)^\dag=C^i_{-p},
\nonumber\\[2pt]
&&
I_m^\dag=I_{-m}, \qquad N_m^\dag=N_{-m}.
\eea
Let us define a primary state $|\psi \rangle$ by
\bea\label{Primary state}
&&
L_0 |\psi \rangle = l_0 |\psi \rangle, \qquad  M_0 |\psi \rangle = m_0 |\psi \rangle,
\nonumber\\[2pt]
&&
I_0 |\psi \rangle = i_0 |\psi \rangle, \qquad  N_0 |\psi \rangle = n_0 |\psi \rangle,
\eea
with $L_m |\psi \rangle=M_m |\psi \rangle=I_m |\psi \rangle=N_m |\psi \rangle=C_p^i |\psi \rangle=0$ for $m>0$ and $p>-\frac12$, where $m_0$, $n_0$, $i_0$ and $l_0$ are positive numbers. Descendants are supposed to
be generated by the integer and half--integer spin operators with negative $m$ and  $p$, respectively. The level of a state is defined with respect to the operator $L_0$ (see \cite{Bagchi:2009pe,Bagchi:2012cy,Campoleoni:2016vsh,Bagchi:2019unf} for the analysis of representations of the $BMS_3$
algebra). There are seven states at the level one, generated by the operators $L_{-1}, M_{-1}, I_{-1}, N_{-1}, C^1_{-\frac12} C^2_{-\frac12}, C^1_{-\frac12} C^1_{-\frac12}$, $C^2_{-\frac12} C^2_{-\frac12}$.
Let us first consider the vacuum state defined by $m_0=n_0=l_0=i_0=0.$ At the level one the only states which give a nonzero contribution to the Kac matrix $K^{(1)}$ are $I_{-1}|\psi \rangle$  and $N_{-1}|\psi \rangle$. In this case the Kac matrix is
   \be
K^{(1)}=\begin{pmatrix}
    0 & 2k\\
    2k & 0
\end{pmatrix}.
\ee
Clearly, to satisfy the semi--positivity condition one needs the vanishing central charge $k=0$. It can also be shown that the vanishing of the central charge is required for a non-vacuum state with nonzero eigenvalues $m_0, n_0, l_0,$ and $i_0$ to have a non-negative norm.
The same issue was encountered in \cite{Castro:2012bc} for the $W^{(2)}_3$ algebra.

To satisfy the unitarity condition with the nonzero central charge,
 we can truncate the algebra  by requiring that the asymptotic fields $\mathcal{N}$ and $\mathcal{I}$ in (\ref{Boundary_Conditions}) are zero, and hence the generators $I_m$ and $N_m$ are removed from the algebra (\ref{Asymptotic_Algebra_Unitarity}).
This also requires to remove the generators $N$ and $I$ from the initial algebra (\ref{Algebra_Lorentz-like}). As a result we have the theory with the vanishing fields $b=v=0$ in (\ref{Action_Extended Einstein}).
To write down the Kac matrix at the level one for the algebra (\ref{Asymptotic_Algebra_Unitarity}) in which $I_m$ and $N_m$ are absent, let us define
a vector of states $|\phi\rangle=(L_{-1}, M_{-1},C^1_{-\frac12} C^2_{-\frac12}, C^1_{-\frac12} C^1_{-\frac12},C^2_{-\frac12} C^2_{-\frac12})|\psi\rangle$.
Using it, the Kac matrix at the level one can be written as $K^{(1)}=|\phi\rangle^\dag\otimes |\phi\rangle$, where the tensor product is for the vector, not for the states. Using the commutation relations (\ref{Asymptotic_Algebra_Unitarity}),
one finds
   \be
K^{(1)}=\begin{pmatrix}
    2l_0 & 2m_0 & i m_0 &0 &0\\
    2m_0 & 0 & 0 &0 &0\\
    -i m_0 & 0 & m_0^2 &0 &0\\
    0 & 0 & 0 &0 & -2m_0^2\\
    0 & 0 & 0 & -2m_0^2 &0
\end{pmatrix}.
\ee
This matrix is positive semi-definite only if the eigenvalue of the operator $M_0$ is vanishing $m_0=0$. The same condition for unitarity was found for the $BMS_3$ algebra in  \cite{Bagchi:2012cy}.
It can be checked that the  positivity condition for the Kac matrix at the $\frac12$-level does not lead to any new restrictions.

To summarize, we have shown that the unitarity condition can only be satisfied for the algebra with the truncated operators $I_m$ and $N_m$.
To answer the question whether there are unitary representations for the truncated algebra one needs to analyze the Kac matrix for every level. We leave it for a future study.

\section{Extending $l$--conformal Galilean algebra}

The Schr\"odinger algebra is a particular instance in the family of nonrelativistic conformal algebras dubbed $l$--conformal Galilean algebras \cite{Negro:1997,Henkel:1997zz}. These algebras and their realizations in physical models have been under extensive study  (see e.g. \cite{Martelli:2009uc,Bagchi:2009my,Duval:2009vt,Duval:2011mi,Fedoruk:2011ua,Galajinsky:2011iz, Galajinsky:2012rp, Andrzejewski:2013ioa,Chernyavsky:2015vav,Chernyavsky:2016mnw,FahadAlshammari:2019att, Galajinsky:2019cfd}.

$l$--conformal Galilean algebras are parameterized by an integer or a half-integer $l$ and $l=\frac12$ corresponds to the Schr\"odinger algebra. We will show that, by analogy with the extension of the  Schr\"odinger algebra, one can also extended the $l$--conformal Galilei algebra with an arbitrary $l$. In what follows we will restrict ourselves to the cases of $d=2$ and $d=1$, where $d$ is the dimension of the Galilean space on which $l$--conformal Galilean algebra naturally acts.
 For our purposes it is convenient to deal with the $l$--conformal Galilean algebra  in the basis considered e.g. in \cite{Martelli:2009uc,Bagchi:2009my}. In the case $d=2$ its nonvanishing commutation relations are
\bea\label{algebra}
&&
[L_m,L_n]=(m-n)L_{m+n},\qquad [L_m,C^i_p ]=(lm-p)C^{i}_{m+p},
\nonumber\\[2pt]
&&
[I, C^i_{p}]=\e^{ij}C_p^j, \qquad m, n=\pm1,0, \quad p=-l, \dots, l, \qquad i=1,2.
\eea
Clearly, the case $d=1$ can be extracted from (\ref{algebra}) by discarding the vector index on the generator $C$ and dropping out the generator $I$.
 As previously, the generators $L_m$ span the conformal subalgebra and $I$ produces rotations in the Galilean space.  $C^i_{-l}$ generate translations, $C^i_{-l+1}$ Galilean boosts, while all the remaining $C^i_m$, $m=-l+2, \dots, l$ are  acceleration generators.
 In the next two subsections we aim to extend the algebra (\ref{algebra}) \footnote{Note that supersymmetric extensions of the $l$-conformal Galilean algebra were constructed earlier in \cite{Galajinsky:2017rls, Galajinsky:2017oiv}.}.

\subsection{Half--integer $l$}
We first construct an extension of the $l$--conformal Galilean algebra for an arbitrary half--integer $l$ in $d=2$. Similar to the structure of the extended Schr\"odinger algebra (\ref{Algebra_Poincare}) and (\ref{Algebra_Infinite-like basis}),  we introduce new generators which appear in the non-zero commutators of the generators $C^i_p$ as follows
\be\label{Commutator}
[C^i_p,C^j_q]=\e^{ij}f^{(l)}_{p,q}M_{p+q}+N_{p,q}^{(l)}\d^{ij},
\ee
where $f^{(l)}_{p,q}$ are symmetric and $N_{p,q}^{(l)}$ are antisymmetric structure constants. We assume that for an arbitrary $l$ there is a Poincar\'e subalgebra (\ref{Algebra_Poincare}) in the extended $l$--conformal Galilei algebra. Hence, the commutation relations above imply that the only nonzero structure constants $f_{p,q}^{(l)}$ are those with $|p+q|\leq 1$. $N^{(l)}_{p,q}$ define a central extension. Their form was found earlier in \cite{Galajinsky:2011iz} in a slightly different notation. The form of $N^{(l)}_{p,q}$ is fixed by the by the Jacobi identity for $(L_m, C_p^i, C^j_q)$
\be
(l m-p)N_{p+m,q}-(l m-q) N_{q+m,p}=0,
\ee
which yields
\be
N^{(l)}_{-p,p}=(-1)^{(p+1/2)^2} \frac{(2l-2p)}{(2l+1)}\prod_{s=\frac12}^p\frac{(2l+2s)}{(2l-2s)} N, \qquad p>0,
\ee
while all the other components of $N^{(l)}_{p,q}$  vanish. By $N$ we denote the only independent element.
The Jacobi identities for $(L_m, C_p^i, C^j_q)$ also require the structure constants $f^{(l)}_{p,q}$ to satisfy the following relation
\be\label{Reccurance}
(m-p-q)f^{(l)}_{p,q}-(l m-p) f^{(l)}_{q,p+m}-(l m-q) f^{(l)}_{p,q+m}=0.
\ee
Curiously, this is exactly the condition on the structure constants in the odd sector of a hyper--Poincar\'e algebra \cite{Fuentealba:2015wza}, a higher-spin generalization of the conventional Poincar\'e supersymmetry algebra first introduced in \cite{Hietarinta:1975fu}.
This restriction implies that the structure constants should satisfy a recurrence relation
 \be\label{Structure constants_Relation}
 f_{p,-p-1}^{(l)}=f_{-p,p+1}^{(l)}=-\frac{p+l+1}{2p}f^{(l)}_{p, -p}, \qquad p>0,
 \ee
 and all the $f^{(l)}_{p, -p}$ can be expressed via
 $f^{(l)}_{-\frac12, \frac12}$
 \cite{Fuentealba:2015wza}
\be\label{Structure constants_Relation_1}
f^{(l)}_{-p,p}=2p \prod_{s=\frac12}^{p-1} \frac{2l+2s+2}{2s-2l} f^{(l)}_{-\frac12,\frac12},
\ee
where we normalize the first element in the recurrence as $f^{(l)}_{-\frac12,\frac12}=1$. There are also two other nontrivial elements of the structure constants which one is not able to identify from (\ref{Structure constants_Relation}). These are   $f^{(l)}_{\frac12,\frac12}=f^{(l)}_{-\frac12,-\frac12}=\frac{1+2l}{2}$.

In \cite{Fuentealba:2015wza} it was pointed out that the structure constants $f^{(l)}_{p,q}$ can be expressed via homogeneous polynomials. In the same manner $N_{p,q}$ can also be presented as polynomials. For instance,  the extended $l=\frac32$ Galilean algebra has the following form
\bea\label{Algebra_3/2}
&&
[L_m, L_n]=(m-n)L_{m+n}, \quad \    \qquad \hspace{0.5mm}[L_m, M_n]=M_{m+n},
\nonumber\\[2pt]
&&
[L_m, C_p^i]=\left(\frac{3 m}{2}-p\right) C^i_{m+p}, \qquad [I, C^{i}_p]=\e^{ij} C^j_p, \qquad m,n=\pm1,0, \quad p,q=\pm\frac32, \pm\frac12,
\nonumber\\[2pt]
&&
[C_p^i, C^j_q]=\frac14 \left(9+8 p q -6p^2-6q^2 \right)\e^{ij} M_{p+q}-\frac{1}{2}  (p-q)\left( p^2+q^2-\frac{5}{2}\right)\d^{ij}N
\eea
The structure constantsin the commutator $[C_p^i, C^j_q]$ are in agreement with the relations  (\ref{Structure constants_Relation}) and (\ref{Structure constants_Relation_1}).
As in the case $l=\frac12$ of the extended Schr\"odinger algebra, we can rewrite the algebra (\ref{Algebra_3/2}) in the $3d$ Lorentz-invariant form (see Appendix B)
\bea\label{Algebra_3/2_Lorentz}
&&
[J^a, J^b]=\e^{abc}J_c, \qquad\qquad\quad\qquad \qquad \hspace{4mm} [J^a, {\mathcal P}^b]=\e^{abc}{\mathcal P}_c,
\nonumber\\[2pt]
&&
[J^a, Z^{b,i}_\a]=\frac32 (Z^{b,i}\g^a)_\a -(Z^{a,i}\gamma^b)_\a, \qquad [I, Z^{a,i}_\a]=\e^{ij} Z^{a,j}_\a,
\nonumber\\[2pt]
&&
[Z^{a,i}_\a, Z^{b,j}_\b]=\e^{ij} \left(-2 (C\g^c)_{\a\b} {\mathcal P}_c\eta^{ab}+\frac52 \e^{abc} C_{\a\b} {\mathcal P}_c+\frac12(C\g^{(a})_{\a\b} {\mathcal P}^{b)}\right)
\nonumber\\[2pt]
&&
\qquad \qquad\qquad\qquad\qquad\qquad\qquad +\d^{ij}\left(\e^{abc} (C\g_c)_{\a\b}-2\eta^{ab} C_{\a\b}\right)N,
\eea
where the higher-spin generator $Z^{ai}_\a$ is gamma-traceless $Z^{a,i}\g_a =0$.

In the case of a generic half-integer $l$ the number of the generators $C_p^i$ are equal to the number of independent components of a symmetric gamma--traceless tensor $Z^{a_1\dots a_{n} i}_\a$ with $n=l-\frac12$. Hence,
by analogy with the hyper--Poincar\'e algebras \cite{Fuentealba:2015wza}, we can present the extended $l$--conformal Galilean algebra in the following form:
\bea\label{Extended_l-conformal Galilei_half-integer}
&&
[J^a, J^b]=\e^{abc}J_c, \qquad [J^a, {\mathcal P}^b]=\e^{abc}{\mathcal P}_c, \qquad [I, Z^{a_1\dots a_n, i}]=\e^{ij} Z^{a_1\dots a_n,j}
\nonumber\\[2pt]
&&
[J^a, Z^{b_1\dots b_n{},i}]=\left(n+\frac12\right)Z^{b_1\dots b_n,i} \gamma^a- Z^{a (b_2\dots b_n|,i} \gamma^{|b_1)},
\nonumber\\[2pt]
&&
[Z^{a_1\dots a_n, i}_\a, Z^{b_1\dots b_n, j}_\b]=\e^{ij}f^{a_1\dots a_n b_1\dots b_n c}_{\a\b} {\mathcal P}_c+\d^{ij} N^{a_1\dots a_n b_1\dots b_n}_{\a\b} N,
\eea
where the structure constants are $SO(1,2)$ invariant tensors
 constructed with the use of the gamma--matrices, Minkowski metric, Levi--Cevita tensor and the charge conjugation matrix.

\subsection{Integer $l$}
For integer $l$ there is no solution for the recurrence relation (\ref{Reccurance}). To resolve this issue we should change the commutation relations of $C^i_p$ in \eqref{Algebra_3/2} as follows
$$[C_p^i, C_q^j]=\d^{ij}f^{(l)}_{p,q} M_{p+q}.$$
W will further restrict ourselves to the case $d=1$ because it is related to $3D$ relativistic systems which is the main topic of this paper.
Then  the commutation relations for the generators $C_p$ have the following form
\be
[C_p, C_q]=f^{(l)}_{p,q} M_{p+q}.
\ee
All the other commutation relations have the same form as in \eqref{Algebra_3/2}, while the generator $I$ is dropped out. The nonzero structure constants $f_{p,q}^{(l)}$ are those with $|p+q|\leq 1$. From the Jacobi identities for the set of generators $(L_m, C_p, C_q)$ we find the following constraint\footnote{Note that for the half-integer $l$ there is no nontrivial solution of (\ref{Jacobi_Integer}).}
 \be\label{Jacobi_Integer}
 (m-p-q)f^{(l)}_{p,q}+(l m-p) f^{(l)}_{q,p+m}-(l m-q) f^{(l)}_{p,q+m}=0.
 \ee
 It implies that non-zero structure constants should be related as in (\ref{Structure constants_Relation}). Explicitly, they are  $f_{0,1}^{(l)}=f_{-1,0}^{(l)}=\frac{l}{2} f_{-1,1}$ and
 \be\label{Structure constants_Relation_2}
 f^{(l)}_{-p,p}=p\prod_{s=1}^{p-1}\frac{l+s+1}{s-l} f^{(l)}_{-1,1}.
 \ee
In what follows, we normalize $f^{(l)}_{-1,1}=1$.
Again, the structure constants have the form of polynomials.

Let us consider two simple examples of the extended $l$--conformal Galilean algebra.

\subsection*{Case $l=1$}
\bea\label{Algebra_l=1}
&&
[L_m, L_n]=(m-n)L_{m+n}, \quad [L_m, M_n]=(m-n)M_{m+n}, \nonumber\\[2pt]
&&
[L_m, C_n]=(m-n)C_{m+n},
\nonumber\\[2pt]
&&
[C_m, C_n]=(m-n)M_{m+n}, \qquad m,n=\pm1,0,
\eea
where we have also redefined $M_m\rightarrow -M_m$ and $C_m\rightarrow 2 C_m$.
This is the Maxwell algebra in three dimensions written in the $BMS_3$-like basis (see e.g. \cite{Concha:2018zeb}). In order to present it in the standard Lorentz-invariant form one makes the redefinition as in eq. (\ref{Redefinition_l=1}) and gets
\bea\label{Algebra_Maxwell}
&&
[J^a,J^b]=\e^{abc}J_c, \qquad\hspace{1mm} [J^a,{\mathcal P}^b]=\e^{abc}{\mathcal P}_c.
\nonumber\\[2pt]
&&
[J^a, Z^b]=\e^{abc}Z_c, \qquad [{\mathcal P}^a,{\mathcal P}^b]=\e^{abc}Z_c,
\eea
which is a conventional form of the $3D$ Maxwell algebra \cite{Bacry:1970ye,Schrader:1972zd}. The generator $Z_{ab}=\e_{abc}Z^c$ of this albebra is associated with a constant electro-magnetic field strength. Note that the role of $Z^a$ and of the translation generator ${\mathcal P}^a$ can be interchanged, and the algebra takes the form of the simplest Hietarinta algebra \cite{Hietarinta:1975fu} used in \cite{Bansal:2018qyz}
\bea\label{Algebra_Maxwell1}
&&
[J^a,J^b]=\e^{abc}J_c, \qquad \hspace{1mm}[J^a,{\mathcal P}^b]=\e^{abc}{\mathcal P}_c.
\nonumber\\[2pt]
&&
[J^a, Z^b]=\e^{abc}Z_c, \qquad [Z^a,Z^b]=\e^{abc}{\mathcal P}_c.
\eea
\subsection*{Case $l=2$}
 \bea\label{Algebra_l=2}
&&
[L_m, L_n]=(m-n)L_{m+n}, \qquad [L_m, M_n]=(m-n)L_{m+n}, \nonumber\\[2pt]
&& [L_m, C_p]=(2m-p)C_{m+p}
\nonumber\\[2pt]
&&
[C_p, C_q]=\frac16 (p-q)(2p^2+2 q^2-p q -8)M_{p+q}, \\[2pt]
&& m,n=\pm1,0, \quad p, q=\pm2, \pm1, 0.\nonumber
\eea
 In a similar fashion we can rewrite $l=2$ commutation relations  (\ref{Algebra_l=2}) in the Lorentz invariant form by redefining the generators as in (\ref{Redefinition_l=2})
  \bea\label{Commutator_l=2}
 &&
[J^a, J^b]=\e^{abc}J_c, \qquad\quad \hspace{2.5mm}[J^a,{\mathcal P}^b]=\e^{abc}{\mathcal P}_c,
\nonumber\\[2pt]
&&
[J^a, Z^{bc}]=\e^{da(b}Z^{c){}_d}, \qquad  [Z^{ab},Z^{cd}]={\mathcal P}_e\e^{ec(a}\eta^{b)d}+{\mathcal P}_e\e^{ed(b}\eta^{a)c},
 \eea
  where the generator $Z^{ab}$ is symmetric and traceless $Z^{ab}\eta_{ab}=0.$

\subsection*{Generic integer $l$}

  As in the case of the half-integer $l$ one can represent the extended integer $l$--conformal Galilean algebra in a $3D$ relativistic form by introducing a higher spin generator $Z^{a_1\dots a_l}$, which is symmetric and traceless. Indeed, the number of generators $C_n$ for a given integer $l$ is equal to $2l+1$, which is exactly the number of independent components of a traceless symmetric tensor of rank $l$ in three dimensions. We thus get the following algebra which is a subclass of the Hietarinta algebras \cite{Hietarinta:1975fu}
    \bea\label{Extended_l-conformal Galilei_integer}
 &&
[J^a, J^b]=\e^{abc}J_c, \qquad\qquad\hspace{13mm} [J^a,{\mathcal P}^b]=\e^{abc}{\mathcal P}_c,
\nonumber\\[2pt]
&&
[J^a, Z^{a_1\dots a_l}]= \e^{ab (a_1}Z^{a_2\dots a_l)}{}_b, \qquad [Z^{a_1\dots a_l}, Z^{b_1\dots b_l}]=f^{a_1\dots a_l b_1\dots b_l c}{\mathcal P}_c,
 \eea
where the structure constants are $SO(1,2)$ invariant tensors respecting the tracelessness of $Z^{a_1\dots a_l}$. These algebras can be further extended by relaxing the traceless condition allowing $Z^{a_1\dots a_l}$ be an arbitrary mixed-symmetry tensor.

\section{Relativistic gravity models with extended $l$--conformal Galilean symmetry}
We shall now construct higher-spin gravity theories which are invariant under local extended $l$-conformal Galilean symmetry.
It can be shown that the $l$-conformal Galilean algebra has a non-degenerate $SO(1,2)$-invariant bilinear form for any $l$. However, instead of exploiting  the standard Chern-Simons construction requiring the explicit use of the bilinear form, we will write down directly the final action and check its symmetry properties starting from the case of the half-integer $l$.

\subsection{Half--integer $l$}
The higher-spin $3D$ gravity action invariant under the local transformations generated by the algebra \eqref{Extended_l-conformal Galilei_half-integer} is \footnote{Its form can be read off from the $l=\frac12$ action (\ref{Action_Extended Einstein}) and also from the hyper--gravity action in \cite{Fuentealba:2015jma,Fuentealba:2015wza}.}
\be\label{ilga}
S=\frac{k}{4\pi}\int_{\mathcal{M}_3}\left( 2e^a R_a-\e^{ij}\bar{\lambda}_{a_1\dots a_n}^{i} \nabla \lambda^{a_1\dots a_n,j}-2v db\right),
\ee
where the covariant derivative is defined by
\be
\nabla \lambda^{a_1\dots a_n,i}=d \lambda^{a_1\dots a_n, i}+\left(n+\frac12\right) \omega^b\gamma^b \lambda^{a_1\dots a_n, i}-\omega^b \gamma^{( a_1} \lambda^{a_2\dots a_n) b,i}-b \e^{ij}\lambda^{a_1\dots a_n,j},
\ee
and $n=l-\frac12$.

For generality, one could also add to the action \eqref{ilga} a Chern-Simons term constructed with the spin connection $\omega^a$ (see e.g. \cite{Salgado:2014jka})
\be\label{CSomega}
S_{\tt m}=\frac{k}{4\pi\,{\tt  m}}\int_{\mathcal{M}_3}\left(\omega^a d\omega_a+\frac 13\epsilon_{abc}\omega^{a}\omega^b\omega^c\right)\,,
\ee
where ${\tt m}$ is the parameter of mass dimension.

Note that the addition of \eqref{CSomega} to \eqref{ilga} does not change the non-dynamical nature of the fields in the bulk, in particular $R^a=0$ on the mass shell, because  $e^a$ and $\omega^a$ are considered as independent fields. This is in contrast to topologically massive gravity \cite{Deser:1982vy,Deser:1981wh} in which the spin connection is {\it a priori} constructed from the dreibein.

By construction, in addition to local Poincar\'e symmetry  this theory enjoys gauge symmetry associated to the generators $Z_\a^{ a_1\dots a_n,i}$, $I$ and $N$ (\ref{Extended_l-conformal Galilei_half-integer}).
Local Poincar\'e transformations read
\bea\label{Gauge Lorentz transformations}
&&
\d e^a=d\a^a+\e^{abc}(e_b\b_c+\omega_b\a_c), \qquad \d\omega^a=d\b^a+\e^{abc}\omega_b \b_c,
\nonumber\\[2pt]
&&
\d \lambda^{a_1\dots a_n,i}=-\left(n+\frac12\right) \b^a\gamma_a \lambda^{a_1\dots a_n,i}+\b_a\gamma^{(a_1}\lambda^{a_2 \dots a_{n}) a,i},
\eea
where $\b^a$ is the gauge parameter, corresponding to the Lorentz rotations $J^a$, while $\a^a$ is the parameter of local translations $\mathcal{P}^a$.
Gauge symmetry transformations generated by $Z$ are given by
\bea
&&
\d e^a=\left(n+\frac12\right)\e^{ij}\bar{\lambda}^{a_1\dots a_n, i} \gamma^a\ve_{a_1\dots a_n}^{j},
\nonumber\\[2pt]
&&
\d \lambda^{a_1\dots a_n,i}=\nabla \ve^{a_1\dots a_n,i}, \qquad \d v=-\bar{\lambda}^{a_1\dots a_n,i}\ve_{a_1\dots a_n}^{i},
\eea
and the gauge parameter is totally symmetric and gamma--traceless.
 For checkin the invariance of the action under these transformations the following identity is useful
\be
\nabla^2\ve^{a_1\dots a_n,i}=\left(n+\frac12\right)R^a\gamma_a \ve^{a_1\dots a_n,i}-R_a\gamma^{(a_1}\ve^{a_2\dots a_n)a,i}-\e^{ij}db\ve^{a_1\dots a_n,j}.
\ee
The gauge transformations associated to the generators $I$ and $N$ are
\be
\d \lambda^{a_1\dots a_n,i}=\kappa \e^{ij} \lambda^{a_1\dots a_n,j}, \qquad \d b=d\kappa, \qquad \d v=d\varphi,
\ee
where $\kappa$ and $\varphi$ are the gauge parameters. The structure of the action \eqref{ilga} is very similar to hypergravity theory  \cite{Aragone:1980rk,Fuentealba:2015jma,Fuentealba:2015wza,Rahman:2019mra}, but it also includes the coupling of the higher-spin fields to the R-Symmetry gauge field $b$.

\subsection{Integer $l$}
Let us now turn to the case of integer $l$. Again, one may see that there exists a bilinear form for the algebra (\ref{Extended_l-conformal Galilei_integer}), but we found it simpler to construct the higher-spin gravity action without using it explicitly. The action (to which one can also add the CS term \eqref{CSomega}) has the following form
\be\label{Action_Integer l}
S=\frac{k}{4\pi}\int_{\mathcal{M}_3} \left(2e^a R_a+\lambda^{a_1\dots a_l}\nabla \lambda_{a_1\dots a_l}\right),
\ee
where the covariant derivative is given by
\be
\nabla \lambda^{a_1\dots a_l}=d \lambda^{a_1\dots a_l}+\e^{bc(a_1} \lambda_b{}^{a_2\dots a_l)}\omega_c.
\ee
In addition to the local Poincar\'e symmetry, which is given by the first row in (\ref{Gauge Lorentz transformations}) and
\be
\d \lambda^{a_1\dots a_l}=-\e^{ab (a_1}\lambda^{a_2\dots a_l)}{}_a \b_b,
\ee
this action is invariant under the gauge transformations
\be
\d e^a=\e^{abc} \lambda_{b b_2\dots b_{l}} \ve_{c}{}^{ b_2\dots b_{l}}, \qquad \d \lambda^{a_1\dots a_l}=\nabla\ve^{a_1\dots a_l},
\ee
associated to the generators $Z^{a_1\ldots a_l}$. In the case $l=1$ the action is invariant under local  symmetry generated by \eqref{Algebra_Maxwell1} which is `dual' to the Maxwell algebra \eqref{Algebra_Maxwell}. A 3D gravity model based on the Hietarinta algebra \eqref{Algebra_Maxwell1} and its higher-spin extensions describing 3D gravity coupled to mixed symmetry fields $\lambda^{a_1\dots a_n}$ were constructed in \cite{Bansal:2018qyz}.
Earlier, the 3D gravity model based on the conventional Maxwell algebra \eqref{Algebra_Maxwell} was constructed and studied in \cite{Salgado:2014jka,Hoseinzadeh:2014bla,Aviles:2018jzw}.\footnote{ Higher-spin extensions of the Maxwell algebra and corresponding gravity models were considered in \cite{Caroca:2017izc}.  See also \cite{Salgado-Rebolledo:2019kft} for a detailed study of the $3D$ Maxwell group, its infinite-dimensional extensions, applications and additional references.}

The most general `bi-gravity' action based on the algebra \eqref{Algebra_Maxwell1} has the following form
\be\label{MH}
S=\frac{k}{4\pi}\int_{\mathcal{M}_3} \,\left[\left(2e^a R_a+\lambda^{a}\nabla \lambda_{a}\right)+ 2{\tt a}\,\lambda^a R_a+\frac 1 {\tt m}\left(\omega^a d\omega_a+\frac 13\epsilon_{abc}\omega^{a}\omega^b\omega^c\right)\right],
\ee
where $T^a=De^a$ and $\tt a$ is a coupling constant parameter in addition to $k$ and $\frac 1 {\tt m}$.

This action is similar to the Maxwell Chern-Simons gravity action of \cite{Salgado:2014jka} based on the  algebra \eqref{Algebra_Maxwell} and can be constructed by using the bilinear form
\be
\langle J_a, P_b\rangle = \eta_{ab}, \qquad \langle Z_a, Z_b\rangle = \eta_{ab},\qquad \langle J_a, J_b\rangle = \frac 1 {\tt m}\eta_{ab}, \qquad \langle J_a, Z_b\rangle = {\tt a}\eta_{ab}.
\ee
To pass from one action to another one should swap the one-form fields $e^a$ with $\lambda^a$.

\subsection{Asymptotic symmetries in $l=\frac32$, $l=1$ and $l=2$ cases}
In this section we will study the asymptotic symmetry of the extended gravity theories described by the actions (\ref{ilga})
and  (\ref{Action_Integer l})  for the cases $l=\frac32$ and $l=1,2$.

\subsubsection*{Case $l=\frac32$}

As in the case $l=\frac12$  of the extended Schr\"odinger algebra discussed in Section \ref{as}, we may relax the boundary conditions and allow for the additional fields to have nonzero excitations, defining the boundary conditions in such a way that they include the $BMS_3$ ones. For simplicity, for the $l=\frac32$ case we assume that the central charge in the algebra (\ref{Algebra_3/2}) is zero and take the boundary conditions in the form
\be\label{boundary condirions l=3/2}
\mathfrak{a}_\phi=\mathfrak{a}_\phi^0+\frac13  \mathcal{C}^iC^i_{-\frac32}, \qquad \mathfrak{a}_t=\mathfrak{a}_t^0,
\ee
where $\mathfrak{a}_\phi^0$ and $\mathfrak{a}_t^0$ are given in (\ref{Boundary_Conditions_BMS}). In order to satisfy the equations of motion we still need the functions $\mathcal{L}$ and $\mathcal{M}$ in (\ref{Boundary_Conditions_BMS}) to be related as in (\ref{Equations of motion_Solution}), while $\mathcal{C}$ should be a function of $\phi$ only. The same restrictions are imposed on the functions describing asymptotic dynamics in the cases $l=1,2$, which we will study below. One may notice a close similarity between these boundary conditions and the ones in the $N=1$ supergravity \cite{Barnich:2014cwa} or in the hypergravity theories \cite{Fuentealba:2015wza}. The algebra-valued parameter $\boldsymbol{\lambda}$, which generates the transformation preserving these boundary conditions, is given by
\bea
&&
\boldsymbol{\lambda}=L_{+1}\ve_L-L_0 \ve_L'+\left(\frac12 \ve_L''-\ve_L \mathcal{M}\right)L_{-1}+\left(\frac12 \ve_M''-\ve_L \mathcal{L}-\ve_M\mathcal{M}-\frac32 \mathcal{C}^i\ve^i\right) M_{-1}
\nonumber\\[2pt]
&&
\qquad+M_{+1}\ve_M-M_0\ve'_M+\e^{ij}\left(C^i_{+\frac32}\ve^j-C^i_{+\frac12}\ve'^j\right)+\e^{ij}C_{-\frac12}^i\left(\frac12\ve''^j-\frac32 \mathcal{M}\ve^j\right)
\nonumber\\[2pt]
&&
\qquad+\e^{ij}C^i_{-\frac32}\left(\frac13 \mathcal{C}^i\ve_L+\frac12 \mathcal{M}'\ve^j+\frac76 \mathcal{M}\ve'^j-\frac16\ve'''^j\right).
\eea
The requirement that the components of the gauge field $\mathfrak{a}_t$ along the time direction be preserved by the same transformation implies that the parameters $\ve_L$ and $\ve_M$ are related as in (\ref{Parameters_Constraint}), while $\ve^i$ is a time independent function. Following the steps of Section \ref{as}, one finds the asymptotic symmetry algebra
\bea\label{asl32}
&&
[L_m, L_n]=i(m-n)L_{m+n}, \qquad [L_m, C^i_p]=i\left(\frac{3m}{2}-p\right) C^i_{m+p},
\nonumber\\[2pt]
&&
[L_m, M_n]=i(m-n)M_{m+n}-i k n^3\d_{m+n,0},
\\[2pt]
&&
[C^i_p, C^j_q]=\left(2pq-\frac32 p^2 -\frac32 q^2\right)\e^{ij} M_{p+q}-\frac{9}{4k}\sum_{s} M_{p+q-s}M_s \e^{ij}-\e^{ij}k p^4 \d_{p+q,0},\nonumber
\eea
where we have only wrote the non-zero commutators.

In contrast to the case $l=\frac12$, the algebra involves a nonlinear term, which is common for asymptotic symmetry algebras of higher-spin gravity theories (see \cite{Campoleoni:2010zq} and references therein).

\subsubsection*{Case $l=1$}
The structure of the asymptotic symmetry of gravity with the gauged Maxwell symmetry \eqref{Algebra_Maxwell} was studied in \cite{Concha:2018zeb}. As we mentioned above, in this case the roles of the  generator $Z_a$ the translation generator $P_a$, and of the corresponding spin-2 fields get interchanged in comparison to the $l=1$ algebra \eqref{Algebra_Maxwell1} and the gravity action \eqref{MH}. As such, in the latter case we have
the different definition of the 3D space-time  and different boundary conditions (see \eqref{Redefinition_l=1} for the redefinition of the generators of \eqref{Algebra_Maxwell1})
\be\label{boundary condirions l=1}
\mathfrak{a}_\phi=\mathfrak{a}_\phi^0-\mathcal{C}(\phi)C_{-1}, \qquad \mathfrak{a}_t=\mathfrak{a}_t^0,
\ee
where $\mathfrak{a}_\phi^0$ and $\mathfrak{a}_t^0$ are given in (\ref{Boundary_Conditions_BMS}).

The corresponding  parameter of the transformations compatible with these boundary conditions are
\bea
&&
\boldsymbol{\lambda}=\left(\frac12 \ve_L''-\ve_L \mathcal{M}\right)L_{-1}+\left(\frac12 \ve_M''-\ve_L \mathcal{L}-\ve_M\mathcal{M}-\mathcal{C}\ve\right) M_{-1}
\\[2pt]
&&
+\left(\frac12 \ve''-\mathcal{M}\ve-\mathcal{C}\ve_L\right)+L_{+1}\ve_L-L_0 \ve_L'+M_{+1}\ve_M-M_0\ve'_M+C_{+1}\ve-C_{0}\ve',\nonumber
\eea
where the parameters $\ve_L$ and $\ve_M$ are related as in (\ref{Parameters_Constraint}) and $\ve$ is time independent.
As a result, we get the asymptotic symmetry algebra similar to that in  \cite{Concha:2018zeb} but with the interchanged role of the generators $M_n$ and $C_n$
\bea\label{asl1}
&&
[L_m, L_n]=i(m-n)L_{m+n}-   \frac {ik} {\tt{m}} n^3 \d_{m+n,0},
\nonumber\\[2pt]
&&
[L_m, C_n]=i\left(m-n\right) C_{m+n}-i {\tt a} k n^3\d_{m+n,0},
\nonumber\\[2pt]
&&
[L_m, M_n]=i(m-n)M_{m+n}-i k n^3\d_{m+n,0},
\nonumber\\[2pt]
&&
[C_m, C_n]=i(m-n)M_{m+n}-i k n^3\d_{m+n,0},
\nonumber\\[2pt]
&&
[M_m,M_n]=0=[M_m,C_n]\,,
\eea
where the central charge in the first line depends on the mass parameter $\tt m$ of the CS spin-connection term and the central charge in the second line is proportional to the coupling constant $\tt a$ associated with the second Einstein-like term in the action \eqref{MH}.

\subsubsection*{Case $l=2$}
Though the theories with integer and half-integer $l$ have different properties and field content, as we have seen previously, they have boundary conditions of a very similar form. For the case $l=2$ we have
\be\label{boundary condirions l=2}
\mathfrak{a}_\phi=\mathfrak{a}_\phi^0+\mathcal{C}(\phi)C_{-2}, \qquad \mathfrak{a}_t=\mathfrak{a}_t^0.
\ee
The parameter of the transformations (\ref{Transformation}) preserving  these boundary conditions has the following form
\bea
&&
\boldsymbol{\lambda}=L_{+1}\ve_L-L_0 \ve_L'+\left(\frac12 \ve_L''-\ve_L \mathcal{M}\right)L_{-1}+\left(\frac12 \ve_M''-\ve_L \mathcal{L}-\ve_M\mathcal{M}+4 \mathcal{C}\ve\right) M_{-1}
\nonumber\\[2pt]
&&
+M_{+1}\ve_M-M_0\ve'_M+C_{+2}\ve-C_{+1}\ve'+C_0\left(\frac12\ve''-2\mathcal{M}\ve\right)
\nonumber\\[2pt]
&&
+C_{-1}\left(-\frac16\ve'''+\frac23 \mathcal{M}'\ve+\frac53 \mathcal{M}\ve'\right)
\nonumber\\[2pt]
&&
+C_{-2}\left(\mathcal{C}\ve_L-\frac16\mathcal{M}''\ve-\frac23 \mathcal{M}\ve''-\frac{7}{12} \mathcal{M}'\ve'+\mathcal{M}^2\ve\right),
\eea
where, again, the parameters $\ve_L$ and $\ve_M$ are related as in (\ref{Parameters_Constraint}) and $\ve = \ve(\phi)$.
The corresponding asymptotic symmetry algebra is
\bea\label{asl2}
&&
[L_m, L_n]=i(m-n)L_{m+n}, \qquad \qquad [L_m, C_p]=i\left(2m-p\right) C_{m+p},
\\[2pt]
&&
[L_m, M_n]=i(m-n)M_{m+n}-i k n^3\d_{m+n,0},
\nonumber\\[2pt]
&&
[C_p, C_q]=(p-q)\left(p q-2p^2-2q^2\right)M_{p+q}-\frac{8}{k}(p-q)\sum_{s} M_{p+q-s}M_s+k q^5 \d_{p+q,0},\nonumber
\eea
which also has the nonlinear term.

As  in the case $l=\frac32$, our boundary conditions for $l=1, 2$ are similar to those in supergravity theories \cite{Barnich:2014cwa, Fuentealba:2015wza}, but with a fermionic generator term  replaced by the bosonic one associated to the generator $C_{-l}$, as  in \eqref{boundary condirions l=2}.
The above consideration can be extended to the case of arbitrary $l$ for which
a suitable choice of boundary conditions should lead to asymptotic symmetries whose algebra is a generalization of \eqref{asl32}, \eqref{asl1} and \eqref{asl2}.
\section{Conclusion}
We have shown that the extended Schr\"odinger algebra and the corresponding Chern-Simons action describing the conformal non-projectable Ho$\breve{\rm r}$ava–Lifshitz gravity constructed in \cite{Hartong:2016yrf}, can be viewed as an extended $3D$ Poincar\'e algebra allowing one to rewrite the Chern-Simons action of \cite{Hartong:2016yrf} in a manifestly $3D$ Lorentz-invariant form. So with a different (relativistic) choice of boundary conditions the same Chern-Simons action describes a relativistic $3D$ theory coupled to two spin-1 gauge fields and a doublet of bosonic spin-3/2 fields.  We have shown that the above theory can be regarded as an asymptotic flat-space contraction (and truncation) of the $SU(1,2)\times SU(1,2)$ Chern-Simons theory with a non principle embedding of $SL(2,\mathbb R)$ into $SU(1,2)$. The asymptotic symmetry algebra of this theory has the form (\ref{Asymptotic_Algebra}). Because of the spin-statistics correspondence for the  spin-$3/2$ fields and the corresponding generators of the gauge symmetry, the asymptotic states on the $2d$ boundary are, in general, not unitary, unless the algebra and the spectrum of states are further truncated.
It would be of interest to analyze a similar issue for the non-relativistic choice of the metric and corresponding boundary conditions associated with the conformal non-projectable Ho$\breve{\rm r}$ava–Lifshitz gravity of \cite{Hartong:2016yrf}.
This study can be carried out following the lines of  \cite{Grumiller:2016pqb,Grumiller:2017sjh} which considered the most general boundary conditions in $3D$ gravity. In this way one may expect to obtaine a centrally extended affine version of the extended Schr\"odinger algebra at the boundary and a corresponding field spectrum describing excitations around the $z=2$ Lifshitz geometries found in \cite{Hartong:2016yrf}.

We have also constructed extensions of $l$-conformal Galiean algebras (with $l=1/2$ referring to the Schr\"odinger algebra) and corresponding relativistic higher-spin gravity theories, and derived their asymptotic symmetries for the cases of $l=\frac32$ and $l=2$. In this regard, it would be of interest to study whether and how these theories can be obtained by an asymptotic flat-space contraction of conventional Chern-Simons higher-spin gravities and their asymptotic  W-algebras. These issues will be considered elsewhere.

\subsection*{Acknowledgements}
The authors are grateful to Andrea Campoleoni, Daniel Grumiller, Jelle Hartong, Radoslav Rashkov, Patricio Salgado-Rebolledo and Evgeny Skvortsov for interest to this work and useful discussions. D.C. would like to thank INFN, Sezione di Padova, for hospitality during initial stages of this work.   D.S. acknowledges support and hospitality extended to him at the ESI (Vienna) Program ``Higher spins and holography'' (March 11-22, 2019) and at the School of Physics and Astrophysics, University of Western Australia where part of this work was done. Work of D.C. was supported by the Russian Science Foundation, grant No 19-11-00005. Work of D.S. was supported in part by the Australian Research Council project No. DP160103633.

\appendix
\numberwithin{equation}{section}

\section{Conventions}
Our conventions are such that the Minkowski metric is given in null coordinates, in which the only nontrivial components of the metric are $\eta^{+-}=\eta^{-+}=\eta^{22}=1.$ Accordingly, the gamma--matrices are given by \be
 \qquad \g^-=\sqrt2 \begin{pmatrix}
0 & 1 \\
0 & 0 \\
\end{pmatrix}, \qquad \g^+=\sqrt2\begin{pmatrix}
0 & 0 \\
 1 & 0 \\
\end{pmatrix}, \qquad \g^2=\begin{pmatrix}
1 &  0 \\
0 & -1 \\
\end{pmatrix},
\ee
and satisfy the identities
\be
\g^{a}\g^b=\eta^{ab}+\e^{abc}\g_c, \qquad (\g_a)^\a{}_\b (\g^a)^\rho{}_\sigma=2\d^\a_\s \d^\rho_\b-\d^\a_\b \d^{\rho}_\s,
\ee
where $\e^{-+2}=1$. We define the conjugate spinor as $\bar\lambda_\a=C_{\a\b}\lambda^\b$, where the conjugation matrix is given by $C_{\a\b}=\e_{\a\b}$ with $\e_{12}=1$. Hence, the conjugation matrix is real and antisymmetric, while its product with a gamma--matrix is symmetric $(C\gamma^a)_{\a\b}=(C\gamma^a)_{\b\a}$.

Throughout the text round brackets denote symmetrization of the indices enclosed by them without a normalization factor, e.g.
\be
\lambda^{(ab)}=\lambda^{ab}+\lambda^{ba}.
\ee

\section{$su(1,2)$ and $sl(3,R)$ algebras}
The commutation relations
\bea\label{Algebra_su(1,2)}
&&
[L_m, L_n]=(m-n)L_{m+n}, \qquad [L_m, W_p]=(2m-p)W_{m+p},
\nonumber\\[2pt]
&&
[W_p, W_q]=\frac{\sigma}{3} (p-q)(2p^2+2q^2-p q-8)L_{p+q},
\eea
with $m,n=\pm1,0$ and $p,q=\pm2,\pm1, 0$, represent the $su(1,2)$ algebra for $\sigma=1$ and $sl(3,R)$ one for $\sigma=-1$. The corresponding bilinear form is
\bea\label{Bilinear_su(1,2)}
&&
\langle L_{-1},L_{+1} \rangle=-1, \qquad \langle L_{0},L_{0} \rangle=\frac12,
\nonumber\\[2pt]
&&
\langle W_{-1},W_{+1} \rangle=\sigma, \qquad \langle W_{-2}, W_{+2} \rangle=-4\sigma, \qquad \langle W_{0},W_{0} \rangle=-\frac{2\sigma}{3}
\eea
The $sl(2,R)$ subalgebra is generated by $(L_{\pm1}, L_0)$ and this embedding of $sl(2,R)$ algebra into $su(1,2)$ is known as principal. The non-principle embedding is obtained by the following choice of the $sl(2,R)$
generators
\bea\label{Redefinition_su(1,2)_1}
\mathcal{L}_{-1}=\frac{\sigma}{4} W_{-2}, \qquad \mathcal{L}_{0}=\frac12 L_{0}, \qquad \mathcal{L}_{+1}=\frac14 W_{+2}.
\eea
And upon the following redefinition of the rest of the generators
\bea
\mathcal{I}=-\frac12 W_0,
\quad
\mathcal{C}^1_{+\frac12}=\frac{\sigma}{2} L_{+1}, \quad \mathcal{C}^2_{+\frac12}=\frac{\sigma}{2} W_{+1}, \quad \mathcal{C}^1_{-\frac12}=\frac12 W_{-1}, \quad \mathcal{C}^2_{-\frac12}=\frac{\sigma}{2} L_{-1},
\eea
one transforms the algebra (\ref{Algebra_su(1,2)}) to the form
\bea\label{cr}
&&
[\mathcal{L}_m,\mathcal{L}_n]=(m-n)\mathcal{L}_{m+n}, \qquad [\mathcal{L}_m, \mathcal{C}_p^i]=\left(\frac{m}{2}-p\right)\mathcal{C}^{i}_{m+p}
\nonumber\\[2pt]
&&
[ \mathcal{C}^i_p, \mathcal{C}^j_q]=\e^{ij} \mathcal{L}_{p+q}-\frac{3}{2}\eta^{ij}(p-q)\mathcal{I}, \qquad [\mathcal{I}, \mathcal{C}^i_p]=\e^{ij} \mathcal{C}^j_{p},
\eea
Here $\eta^{ij}=diag(\sigma, 1)$ and the summation over the indices $(i,j)$ is performed with respect to this metric. The difference between the $sl(3,R)$ and $su(1,2)$ algebra is that  in $sl(3,R)$ the generator $\mathcal I$ is associated with a non-compact $so(1,1)$ subalgebra, while in $su(1,2)$ it generates $so(2)$ rotations. For $\sigma=1$
the commutation relations \eqref{cr} defining $su(1,2)$ can be written in the form (\ref{Algebra_AdS_l=1/2}) upon the redefinition
\bea\label{Redefinition_su(1,2)_2}
&&
\mathcal{J}^-=-\frac{1}{\sqrt2}\mathcal{L}_{-1}, \qquad \mathcal{J}^+=\frac{1}{\sqrt2}\mathcal{L}_{+1}, \qquad \mathcal{J}^2=\mathcal{L}_0,
\nonumber\\[2pt]
&&
\mathcal{Z}^i_1=\frac{1}{\sqrt2} \mathcal{C}^i_{-\frac12}, \qquad \mathcal{Z}^i_2=\frac{1}{\sqrt2} \mathcal{C}^i_{+\frac12}.
\eea

\section{$3D$ Lorentz-covariant form of the extended $l$-conformal Galilean algebra}
Here we list the redefinition of the generators which bring the commutation relations of the extended $l$-conformal Galilean algebra (for $l=1, \frac32$  and 2) to the Lorentz-covariant form. For the each case the redefinition of the generators of the conformal subalgebra is
\bea
\sqrt2 J^-=-L_{-1}, \quad \sqrt2 J^+=L_{+1}, \quad J^2=L_0,
\eea
Redefinitions of all the other generators are:

$\bullet$ $l=1$, from (\ref{Algebra_l=1}) to (\ref{Algebra_Maxwell})
\bea\label{Redefinition_l=1}
&&
{\mathcal P}^-=-\sqrt2 M_{-1}, \quad {\mathcal P}^+=\sqrt2 M_{+1}, \quad {\mathcal P}^2=2 M_0,
\nonumber\\[2pt]
&&
Z^-=C_{-1}, \quad Z^+=-C_{+1}, \quad Z^2=-\sqrt2 C_0.
\eea
\\
$\bullet$ $l=\frac32$,  from (\ref{Algebra_3/2}) to (\ref{Algebra_3/2_Lorentz})
\bea
&&
{\mathcal P}^-=-\sqrt2 M_{-1}, \qquad {\mathcal P}^+=\sqrt2 M_{+1}, \qquad {\mathcal P}^2=2 M_{0},
\nonumber\\[2pt]
&&
Z^{-,i}_1=C_{-\frac32}^i, \qquad Z^{-,i}_2=C_{-\frac12}^i, \qquad Z^{+,i}_1=-C_{+\frac12}, \qquad Z^{+,i}_2=-C_{+\frac32}.
\eea
Note also that the condition $(Z^{a,i}\g_a )^\a=0$ implies that $\sqrt2 Z^{+,i}_1=Z^{2,i}_2$ and $-\sqrt2 Z^{-,i}_2=Z^{2,i}_1$.
\\

$\bullet$ $l=2$, from (\ref{Algebra_l=2}) to (\ref{Commutator_l=2})
\bea\label{Redefinition_l=2}
&&
Z^{--}=C_{-2}, \quad Z^{-+}=- C_{0}, \quad Z^{-2}=-\sqrt2 C_{-1}, \quad Z^{++}= C_2, \quad Z^{+2}=\sqrt{2}C_1,
\nonumber\\[2pt]
&&
{\mathcal P}^-=-\sqrt2 M_{-1},\quad {\mathcal P}^+=\sqrt2 M_{+1}, \quad {\mathcal P}^2=2 M_0.
\eea
The traceless condition $Z^{ab}\eta_{ab}=0$ implies $Z^{22}=-2 Z^{-+}$.

\section{Contraction of $W^{(2)}_{1,2}\oplus W^{(2)}_{1,2}$ algebra}

The contraction of the $W^{(2)}_{3}\oplus W^{(2)}_{3}$ was considered in \cite{Riegler:2016hah}.
It is reasonable to expect that the contraction of the direct product of two $W^{(2)}_{1,2}$ algebras  with the finite-dimensional subalgebra $su(1,2)$ leads to the asymptotic symmetry algebra (\ref{Asymptotic_Algebra}).
 The $W^{(2)}_{1,2}$  has the following form
\bea
&&
[\mathcal{L}_m,\mathcal{L}_n]=(m-n)\mathcal{L}_{m+n}-\frac{\rho}{2}k n^3\d_{m+n,0},
\nonumber\\[2pt]
&&
[\mathcal{L}_m,\mathcal{I}_n]=-n \mathcal{I}_{m+n},
\nonumber\\[2pt]
&&
[\mathcal{I}_m, \mathcal{I}_n]=\frac{2}{3}k \rho m \d_{m+n,0},
\nonumber\\[2pt]
&&
[\mathcal{L}_m, \mathcal{C}^i_p]=\left(\frac{m}{2}-p\right) \mathcal{C}^i_{m+p},
\nonumber\\[2pt]
&&
[\mathcal{I}_m, \mathcal{C}^i_p]=-\e^{ij}\mathcal{C}^j_{m+p},
\\[2pt]
&&
[\mathcal{C}^i_p, \mathcal{C}^j_q]=-\e^{ij} \left(\mathcal{L}_{p+q}-\frac{3}{k \r}\sum_{s} \mathcal{I}_{p+q-s} \mathcal{I}_s+k\rho p^2\d_{p+q,0}\right)-\frac{3}{2}\d^{ij} (p-q) \mathcal{I}_{p+q},\nonumber
\eea
where $\rho$ is a parameter proportional to the $AdS_3$ radius.

 Let us take two copies of the algebra differed by $\pm$ superscript and define
\bea
&&
L_m=i(\mathcal{L}^+_m-\mathcal{L}^-_{-m}), \qquad M_m=\frac{1}{\r}(\mathcal{L}^+_m+\mathcal{L}^-_{-m}), \qquad C^i_p=\sqrt\frac{2}{\r}\mathcal{C}^{+,i}_p,
\nonumber\\[2pt]
&&
I_m=\mathcal{I}^+_m-\mathcal{I}^-_{-m}, \qquad N_m=\frac{2}{3}\frac{i}{\r} (\mathcal{I}^+_m+\mathcal{I}^-_{-m}).
\eea
Taking the limit $\r\rightarrow\infty$ and truncating the generators $C^{-,i}_p$ one finds the algebra of the form (\ref{Asymptotic_Algebra}) except for the commutators
\bea
&&
[M_m, I_n]=\frac{2}{3} i n N_{m+n},
\\[2pt]
&&
[C_p^i, C_q^j]=-\e^{ij}\left(M_{p+q}+\frac{3}{2k}\sum_s N_{p+q-s}N_s+2 k q^2 \d_{p+q,0}\right)+i(p-q) N_{p+q} \d^{ij}.\nonumber
\eea
Making the redefinition $M_m\rightarrow M_m-\frac{3}{2k}\sum_s N_{m-s}N_s$ these commutation relations take the form of (\ref{Asymptotic_Algebra}). In particular, note that this redefinition keeps intact the form of the commutator $[L_m, M_n]$ in (\ref{Asymptotic_Algebra}).

\if{}
\bibliographystyle{abe}
\bibliography{references}{}
\end{document}
\fi{}

\providecommand{\href}[2]{#2}\begingroup\raggedright\endgroup

\end{document}